\documentstyle[psfig,12pt]{article}

\makeatletter

\def\@aabuffer{}
\def\author #1 {\expandafter\def\expandafter\@aabuffer\expandafter
{\@aabuffer \small\rm  #1\relax \par\vspace{0.5em}}}
\def\address#1{\expandafter\def\expandafter\@aabuffer\expandafter
{\@aabuffer \small\it #1\relax \par\vspace{1em}}}

\def\maketitle{
\begin{center}
   {\baselineskip 14pt \bf \@title \par}
   \vskip 2em			   % Vertical space after title.
   \@aabuffer\relax
\end{center} \par
\gdef\@aabuffer{}
}

\def\abstracts#1{
\begin{center}
{\begin{minipage}{4.2truein}
		 \footnotesize
		 \parindent=0pt #1\par
		 \end{minipage}}\end{center}
		 \vskip 2em \par}

\fussy
\def\be{\begin{equation}}
\def\ee{\end{equation}}

\def\section{\@startsection {section}{1}{\z@}{-3.25ex plus -1ex minus
    -.2ex}{1.5ex plus .2ex}{\bf }}

\def\subsection{\@startsection{subsection}{2}{\z@}{-3.25ex plus -1ex minus
    -.2ex}{1.5ex plus .2ex}{\it }}

\renewenvironment{thebibliography}[1]
	{\begin{list}{\arabic{enumi}.}
	{\usecounter{enumi}\setlength{\parsep}{0pt}
	 \setlength{\itemsep}{0pt}
	 \settowidth
	{\labelwidth}{#1.}\sloppy}}{\end{list}}

\def\makecaption#1#2{\vskip15pt  \noindent 
{\small #1: #2}}

\begin{document}

\leftline{\footnotesize Continuum Models and Discrete Systems (CMDS9)}
\leftline{\footnotesize Proceedings of the 9th International
Symposium}
\leftline{\footnotesize June 29--July 3, 1998, Istanbul, Turkey.
Editors E. Inan \& K. Z. Markov}
\leftline{\footnotesize \copyright\kern2pt 1999 World Scientific
Publishing
Co., pp.~00--00}
\vspace*{0.5cm}

\def\newblock{}
\def\dend#1{{\if*#1{\it Paenibacillus dendritiformis}\else
                {\it P. dendritiformis}\fi}}
\def\Tvar{var. {\it dendron}}
\def\Cvar{var. {\it chiralis}}
\def\Tname#1{{\if*#1\dend* \Tvar\else
                \if-#1\dend{} \Tvar\else
                 \dend{} \Tvar{} #1\fi\fi}}
\def\Cname#1{{\if*#1\dend* \Cvar\else
                \if-#1\dend{} \Cvar\else
                 \dend{} \Cvar{} #1\fi\fi}}
\def\Vname#1{{\if*#1\eddi{Paenibacillus}{V}\else
                \if-#1\eddi{P.}{V}\else \eddi{P.}{V} #1\fi\fi}}
\def\bsub#1{{\if*#1{\it Bacillus subtilis}\else
                \if-#1{\it B. subtilis}\else {\it B. subtilis}
#1\fi\fi}}
\def\bacil#1{\if *#1{Bacillus}\else{B.}\fi}
\def\bcirc#1{\if *#1{\it \bacil* circulans}\else
                \if -#1{\it \bacil{} circulans}\else
                   {\it \bacil{} circulans} #1\fi\fi}
\def\ecoli#1{{\if*#1{\it Escherichia coli}\else
                \if-#1{\it E. coli}\else {\it E. coli} #1\fi\fi}}
\def\salmon#1{{\if*#1{\it Salmonella typhimurium}\else
                \if-#1{\it S. typhimurium}\else
                {\it S. typhimurium} #1\fi\fi}}
\def\prot{{\it Proteus mirabilis }}
\def\myxo#1{{\if*#1{\it Myxococcus xanthus}\else
                \if-#1{\it M. xanthus}\else
                {\it M. xanthus} #1\fi\fi}}
\def\T{{${\cal T }$} }
\def\C{{${\cal C }$} }
\def\Tm{{\T morphotype }}
\def\TM{{\T Morphotype }}
\def\Tme#1{{\T morphotype#1}}
\def\equ#1{{Eq.~(\ref{#1})}}
\def\partderiv#1#2{{\partial #1\over\partial #2}}
\def\deriv#1#2{{d #1\over d #2}}
\def\etal{{\it et al. }}

\title{CONTINUOUS AND DISCRETE MODELS OF COOPERATION IN COMPLEX
BACTERIAL COLONIES}

\author{I. COHEN, I. GOLDING, Y. KOZLOVSKY and E.  BEN-JACOB}

\address{School of Physics and Astronomy, Raymond and Beverly Sackler
  Faculty of Exact Sciences, Tel Aviv University, \\
Tel Aviv 69978, Israel}

\maketitle

\abstracts{
We study the effect of discreteness on various models for patterning
in bacterial colonies. In a bacterial colony with branching pattern,
there are discrete entities -- bacteria -- which are only two orders
of magnitude smaller than the elements of the macroscopic pattern. We
present two types of models. The first is the Communicating Walkers
model, a hybrid model composed of both continuous fields and discrete
entities -- walkers, which are coarse-graining of the bacteria.
Models of the second type are systems of reaction diffusion
equations, where the branching of the pattern is due to non-constant
diffusion coefficient of the bacterial field. The diffusion
coefficient represents the effect of self-generated lubrication fluid
on the bacterial movement. We implement the discreteness of the
biological system by introducing a cutoff in the growth term at low
bacterial densities. We demonstrate that the cutoff does not improve
the models in any way. Its only effect is to decrease the effective
surface tension of the front, making it more sensitive to anisotropy.
We compare the models by introducing food chemotaxis and repulsive
chemotactic signaling into the models. We find that the growth
dynamics of the Communication Walkers model and the growth dynamics
of the Non-Linear diffusion model are affected in the same manner.
From such similarities and from the insensitivity of the
Communication Walkers model to implicit anisotropy we conclude that
the increased discreteness, introduced be the coarse-graining of the
walkers, is small enough to be neglected.
 \\[2pt]
%% {\bf 1991/95 Math.\ Subject Class.} 35K57, 92B05, 92C15
}

\section{Introduction} 

The endless array of patterns and shapes in nature has long been a 
source of joy and wonder to layman and scientists alike \cite 
{Kepler,Mandelbrot77b,Vicsek89,BenJacob93,BenJacob97}. During the last
decade, there were exciting developments in the understanding of
pattern determination in non-living systems
\cite{KKL88,Langer89,BG90,BenJacob93}. The attention of many
researchers is now shifting towards living systems, in a hope to use
these new insights for the study of pattern forming processes in living
systems \cite[and references therein]{BCL98}. Bacterial colonies offer
a suitable subject for such research. In some senses they are similar
enough to non-living systems so as their study can benefit from the
knowledge about non-living systems, yet their building blocks
(bacteria) are complex enough to ensure ever so new surprises.

In figure 1 we show representative branching patterns of bacterial  
colonies. These colonies are made up of about $10^{10}$ bacteria of  
the type \Tname* (see \cite{BSST92,BTSA94} for first reference in the  
literature and \cite {TBG98} for identification). For other studies of
branching bacterial patterns see \cite{FM89,MF90,MKNIHY92,MM95,MS96}.
Each colony is  grown in a standard petri-dish on a thin layer of agar
(semi-solid  jelly). The bacteria cannot move on the dry surface and
cooperatively  
they produce a layer of lubrication fluid in which they swim (Fig.  
2). Bacterial swimming is a random-walk-like movement, in which the  
bacteria propel themselves in nearly straight runs separated by brief  
tumbling. The bacteria consume nutrient from the media, nutrient  
which are given in limited supply. The growth of a colony is limited  
by the diffusion of nutrients towards the colony -- the bacterial  
reproduction rate that determines the growth rate of the colony is  
limited by the level of nutrients available for the cells. Note,  
however, that a single bacterium put alone on the agar can reproduce,  
grow in numbers and make a new colony. 

Bacterial colonies entangle entities in many length scales: the  
colony as a whole is the range of several $cm$; the individual  
branches are of width in the range of $mm$ and less; the individual  
bacteria are in the range of $\mu m$, so is the width of the colony's  
boundary; and chemicals in the agar such as the constitutes of the  
nutrient are on the molecular length scale. 

Kessler and Levine \cite{KL98} studied discrete pattern forming  
systems, using reaction-diffusion models with linear diffusion and  
various growth terms. They showed that the ability of the system to  
form two-dimensional patterns depend on the derivative of the growth  
term (reaction term) at zero densities. With a negative derivative,  
the system can form branching pattern; with a positive derivative,  
the system can form only compact patterns with circular envelope.  
They accounted for the discreteness of the system by introducing a  
low-densities-cutoff in the growth term. Doing so to a growth term  
with positive derivative at zero can introduce bumps to the pattern,  
which is a manifestation of a diffusive instability in the
two-dimensional front (the first step towards branching pattern). 

We present here three models for growth of the bacterial colonies.  
The first is the Communicating Walkers model (Sec.  
\ref{sec:CW:model}) which includes discrete entities to describe the  
bacteria, continuous fields to describe chemicals in the agar and an  
explicit free boundary for the colony's edge. The second model is a  
continuous one, a reaction-diffusion model that couples the bacterial  
movement to a field of lubrication fluid (Sec.  
\ref{sec:lubrication}). The diffusion coefficients of the bacterial  
field and the lubrication field depend on the lubrication fluid,  
resulting in a spontaneous formation of a sharp boundary to the  
colony. The third model tries to simplify the former model and  
dispose of the lubrication field by introducing a density-dependent  
diffusion of the bacterial field (Sec. \ref {nonlinear}). We discuss  
the effect of a cutoff in the growth term in the two continuous  
models. We than turn our attention to various features of the  
observed bacterial patterns and see similarities in the different  
models' ability to reproduce these features (Sec. \ref{sec:chemo}). 

\section{The Communicating Walkers Model : An Hybrid Model} 

\label{sec:CW:model} 

The Communicating Walkers model \cite{BSTCCV94a} was inspired by the 
diffusion-transition scheme used to study solidification from 
supersaturated solutions \cite{SKBLM92a,SKBLM92b,Shochet95}. The 
former is a hybridization of the ``continuous'' and ``atomistic'' 
approaches used in the study of non-living systems. The diffusion of 
the chemicals is handled by solving a continuous diffusion equation 
(including sources and sinks) on a tridiagonal lattice with a lattice 
constant $a_0$. The bacterial cells are represented by walkers 
allowing a more detailed description. In a typical experiment there 
are $10^9-10^{10}$ cells in a petri-dish at the end of the growth. 
Hence it is impractical to incorporate into the model each and every 
cell. Instead, each of the walkers represents about $10^4-10^5$ cells 
so that we work with $10^4-10^6$ walkers in one numerical 
``experiment''. 

The walkers perform an off-lattice random walk on a plane within an 
envelope representing the boundary of the wetting fluid. This 
envelope is defined on the same triangular lattice where the 
diffusion equations are solved. To incorporate the swimming of the 
bacteria into the model, at each time step each of the active walkers 
(motile and metabolizing, as described below) moves a step of size 
$d<a_0$ at a random angle $\Theta $. Starting from location 
$\vec{r_i}$, it attempts to move to a new location $\vec{ 
r^{\prime}_i}$ given by:  
\begin{equation} 
\vec{r^{\prime}_i} = \vec{r_i} + d(cos \Theta ; sin \Theta)~. 
\end{equation} 
If $\vec {r^{\prime}_i}$ is outside the envelope, the walker does not 
move. A counter on the segment of the envelope which would have been 
crossed by the movement $\vec{r_i}\rightarrow\vec{r^{\prime}_i}$ is 
increased by one. When the segment counter reaches a specified number 
of hits $N_c $, the envelope propagates one lattice step and an 
additional lattice cell is added to the colony. This requirement of 
$N_c $ hits represent the colony propagation through wetting of 
unoccupied areas by the bacteria. Note that $ N_c $ is related to the 
agar dryness, as more wetting fluid must be produced (more 
``collisions'' are needed) to push the envelope on a harder 
substrate. 

Motivated by the presence of a maximal growth rate of the bacteria 
even for optimal conditions, each walker in the model consumes food 
at a constant rate $\Omega_c $ if sufficient food is available.
We represent the metabolic state of the $i$-th walker by an 'internal
energy' $E_i $. The rate of change of the internal energy is given by
\begin{equation} 
{\frac{d E_i}{d t }} = \kappa C_{consumed} - {\frac{E_m }{\tau_R}}~, 
\end{equation} 
where $\kappa $ is a conversion factor from food to internal energy 
($\kappa \cong 5\cdot 10^3 cal/g$) and $E_m $ represent the total 
energy loss for all processes over the reproduction time $\tau_R$, 
excluding energy loss for cell division. $C_{consumed}$ is  
$ 
C_{consumed} \equiv \min \left( \Omega_C , \Omega_C^{\prime}\right)~, 
$ 
where $\Omega_C^{\prime}$ is the maximal rate of food consumption as 
limited by the locally available food \cite{BCC95b}. 
When sufficient food is available, $E_i$ increases until it reaches a 
threshold energy. Upon reaching this threshold, the walker divides 
into two. When a walker is starved for long interval of time, $E_i$ 
drops to zero and the walker ``freezes''. This ``freezing'' 
represents entering a pre-spore state (starting the process of 
sporulation, see section \ref{sec:repulsive}). 

We represent the diffusion of nutrients by solving the diffusion 
equation for a single agent whose concentration is denoted by 
$n(\vec{r},t)$:  
\begin{equation} 
{\frac{\partial n}{\partial t}}=D_n \nabla^2C - b C_{consumed}~, 
\end{equation} 
where the last term includes the consumption of food by the walkers 
($b $ is their density). The equation is solved on the tridiagonal 
lattice. The simulations are started with inoculum of walkers at the 
center and a uniform distribution of the nutrient. 

Results of numerical simulations of the model are shown in figure 3. 
As in the case of real bacterial colonies, the patterns are compact 
at high nutrient levels and become fractal with decreasing food 
level. For a given nutrient level, the patterns are more ramified as 
the agar concentration increases. The results shown in figure 3 do 
capture some features of the experimentally observed patterns. 
However, at this stage the model does not account for some critical 
features, such as the ability of the bacteria to develop organized 
patterns at very low nutrient levels. Ben-Jacob \etal 
\cite{BCC96,CCB96,BC97,BenJacob97} suggested that chemotactic 
signaling must be included in the model to produce these features 
(see section \ref{sec:chemo}). 

\section{A Layer of Lubrication} 

\label{sec:lubrication} 

The Lubricating Bacteria model is a reaction-diffusion model for the 
bacterial colonies \cite{GKCB98}. This model includes four coupled 
fields. One field describes the bacterial density $b(\vec{x},t)$, the 
second describe the height of lubrication layer in which the bacteria 
swim $l(\vec{x },t)$, third field describes the nutrients 
$n(\vec{x},t)$ and the fourth field is the stationary bacteria that 
``freeze'' and begin to sporulate $s( \vec{x},t)$ (see section 
\ref{sec:repulsive}). 

The dynamics of the bacterial field $b$ consists of two parts; 
a diffusion term which is coupled to the lubrication field and a 
reaction part which contains terms for reproduction and death. 
Following the same arguments presented for the Communicating Walkers 
model, we get a reaction term of the form $\left( \kappa b\min 
(\Omega _C,n)- E_m b/\tau _R\right) $. Assuming that nutrient is 
always the factor limiting the bacterial growth we get, upon 
rescaling, the growth term $bn-\mu b$ ($\mu $ constant). 

We now turn to the bacterial movement. In a uniform layer of liquid, 
bacterial swimming is a random walk with variable step length and can 
be approximated by diffusion. The layer of lubrication is not 
uniform, and its height affects the bacterial movement. An increase 
in the amount of lubrication decreases the friction between the 
bacteria and the agar surface. The term 'friction' is used here in a 
very loose manner to represent the total effect of any force or 
process that slows down the bacteria. 
As the bacterial motion is over-damped, the local speed of the 
bacteria is proportional to the self-generated propulsion force 
divided by the friction. It can be shown that variation of the speed 
leads to variation of the diffusion coefficient, with the diffusion 
coefficient proportional to the speed to the power of two. We assume 
that the friction is inversely related to the local lubrication 
height through some power law: friction$\sim l^\gamma $ and $\gamma 
<0$. The bacterial flux is:  
\begin{equation} 
\vec{J_b}=-D_bl^{-2\gamma }\nabla b 
\end{equation} 

The lubrication field $l$ is the local height of the lubrication 
fluid on the agar surface. Its dynamics is given by:  
\begin{equation} 
{\frac{\partial l}{\partial t}}=-\nabla \cdot \vec{J_l}+\Gamma b 
n(l_{max}-l) - \lambda l 
\end{equation} 
where $\vec{J_l}$ is the fluid flux (to be discussed), $\Gamma $ is 
the production rate and $\lambda $ is the absorption rate of the 
fluid by the agar. $\lambda$ is inversely related to the agar 
dryness. 

The fluid production is assumed to depend on the bacterial density. 
As the production of lubrication probably demands substantial energy, 
it also depends on the nutrient's level. We assume the simplest case 
where the production depends linearly on the concentrations of both 
the bacteria and the nutrients. 

The lubrication fluid flows by diffusion and by convection caused by 
bacterial motion. A simple description of the convection is that as 
each bacterium moves, it drags along with it the fluid surrounding 
it.  
\begin{equation} 
\vec{J_l}=-D_ll^\eta \nabla l+j\vec{J_b} 
\end{equation} 
where $D_l$ is a lubrication diffusion constant, $\vec{J_b}$ is the 
bacterial flux and $j$ is the amount of fluid dragged by each 
bacterium. The diffusion term of the fluid depends on the height of 
the fluid to the power $ \eta >0$ (the nonlinearity in the diffusion 
of the lubrication, a very complex fluid, is motivated by 
hydrodynamics of simple fluids). The nonlinearity causes the fluid to 
have a sharp boundary at the front of the colony, as is observed in 
the bacterial colonies (Fig. 4). 

The complete model for the bacterial colony is:  
\begin{eqnarray} 
{\frac{\partial b}{\partial t}} &=&D_b\nabla \cdot (l^{-2\gamma  
}\nabla b)+bn-\mu b \\ 
{\frac{\partial n}{\partial t}} &=&D_n\nabla ^2n-b n \nonumber \\ 
{\frac{\partial l}{\partial t}} &=&\nabla \cdot (D_ll^\eta \nabla 
l+jD_bl^{-2\gamma }\nabla b)+\Gamma bn(l_{max}-l)-\lambda l \nonumber  
\\ 
{\frac{\partial s}{\partial t}} &=&\mu b \nonumber 
\end{eqnarray} 
The second term in the equation for $b$ represents the reproduction 
of the bacteria. The reproduction depends on the local amount of 
nutrient and it reduces this amount. The third term in the equation 
for $b$ represents the process of bacterial ``freezing''. For the 
initial condition, we set $n$ to have uniform distribution of level 
$n_0$, $b$ to have compact support at the center, and the other 
fields to be zero everywhere. 

Preliminary results show that the model can reproduce branching 
patterns, similar to the bacterial colonies (Fig. 5). At low values 
of absorption rate, the model exhibits dense fingers. At higher 
absorption rates the model exhibits finer branches. We also obtain 
finer branches if we change other parameters that effectively 
decrease the amount of lubrication. We can relate these conditions to 
high agar concentration. 

We can now check the effect of bacterial discreteness on the observed 
colonial patterns. Following Kessler and Levine \cite{KL98}, we 
introduce the discreteness of the system into the continuous model by 
repressing the growth term at low bacterial densities (``half a 
bacterium cannot reproduce''). The growth term is multiplied by a 
Heaviside step function $ \Theta (b-\beta )$, where $\beta $ is the 
threshold density for growth. In figure 6 we show the effect of 
various values of $\beta $ on the pattern. High cutoff values make 
the model more sensitive to the implicit anisotropy of the underlying 
tridiagonal lattice used in the simulation. The result is dendritic 
growth with marked 6-fold symmetry of the pattern. Increased values 
of cutoff also decrease the maximal values of $b$ reached in the 
simulations (and the total area occupied by the colony). 

The reason for the pattern turning dendritic is as follows: the 
difference between tip-splitting growth and dendritic growth is the 
relative strength of the effect of anisotropy and an effective 
surface tension \cite {BenJacob93}. In the Lubricating Bacteria model 
there is no explicit anisotropy and no explicit surface tension. The 
implicit anisotropy is related to the underlying lattice, and the 
effective surface tension is related to the width of the front. The 
cutoff prevents the growth at the outer parts of the front, thus 
making it thinner, reduces the effective surface tension and enables 
the implicit anisotropy to express itself. 

We stress that it is possible to find a range of parameters in which 
the growth patterns resemble the bacterial patterns, in spite a high 
value of cutoff. Yet the cutoff does not improve the model in any 
sense, it introduces an additional parameter, and it slows the 
numerical simulation. We believe that the well-defined boundary makes 
the cutoff (as a representation of the bacterial discreteness) 
unnecessary. 

\section{Non-Linear Diffusion}
\label{nonlinear}

It is possible to introduce a
simplified model, where the fluid field is not included, and is replaced
 by a density-dependent
diffusion coefficient for the bacteria $D_b \sim b^k$ \cite{Cohen97,BCL98}.
Such a term can be justified by a few assumptions 
 about the dynamics at low
bacterial and low lubrication density:\\
--  The production of lubricant is proportional to the bacterial
  density to the power $\alpha >0$ ($\alpha=1$ in the previous mode).\\
-- There is a sink in the equation for the time evolution of the
  lubrication field, e.g. absorption of the lubricant into the agar.
  This sink is proportional to the lubrication density to the power
  $\beta >0$ ($\beta=1$ in the previous mode).\\
--  Over the bacterial length scale, the two processes above are
  much faster than the diffusion process, so the lubrication density
  is proportional to the bacterial density to the power of $\beta /
  \alpha$.\\
-- The friction is proportional to the lubrication density to the
  power $\gamma < 0$.\\
Given the above assumptions, the lubrication field can be removed from
the dynamics and be replaced by a density dependent diffusion
coefficient. This coefficient is proportional to the bacterial density
to the power $k \equiv -2 \gamma \beta / \alpha > 0$

A model of this type is offered by Kitsunezaki\cite{Kitsunezaki97}:
 \begin{eqnarray}
\label{kitsunezaki-eqn}
\lefteqn{ \partderiv{b}{t} = \nabla (D_0 b^k \nabla
  b) + n b - \mu b}\\
\lefteqn{ \partderiv{n}{t} = \nabla ^2 n - b n}\\
\lefteqn{ \partderiv{s}{t} = \mu b}
\end{eqnarray}
For $k>0$ the 1D model gives rise to a front ``wall'',
with compact support (i.e. $b=0$ outside a finite domain). For
$k>1$ this wall has an infinite slope.
The model exhibits branching patterns for suitable parameter values
and initial conditions, as depicted in Fig. 7.
Increasing the initial nutrient level makes the colonies more
dense, similarly to what happens in the other models.

As in the Lubricating Bacteria model, adding the ``Kessler and Levine
correction'' to the model,
i.e. making the growth term disappear for $b<\beta$, does not
seem to make the patterns ``better'', or closer to the experimental
observations (Fig. 8).
The apparent increased sensitivity to the implicit anisotropy results
from the narrowed front, which decreases the effective surface tension.

\section{Chemotaxis}

\label{sec:chemo}

So far, we have tested the models for they ability to reproduce
macroscopic patterns and microscopic dynamics of the bacterial
colonies. All succeeded equally well, reproducing some aspects of the
microscopic dynamics and the patterns in some range of nutrient level
and agar concentration, but so can do other models \cite[and
reference there in]{GKCB98}. We will now extend the Communicating
Walkers model and the Non-Linear Diffusion model to test for their
success in describing other aspects of the bacterial colonies
involving chemotaxis and chemotactic signaling (which are believed to
by used by the bacteria \cite{BCC96,CCB96,BC97,BenJacob97}).
Chemotaxis means changes in the movement of the cell in response to a
gradient of certain chemical field
\cite{Adler69,BP77,Lacki81,Berg93}. The movement is biased along the
gradient either in the gradient direction or in the opposite
direction. Usually chemotactic response means a response to an
externally produced field, like in the case of chemotaxis towards
food. However, the chemotactic response can be also to a field
produced directly or indirectly by the bacterial cells. We will refer
to this case as chemotactic signaling.
The bacteria sense the local concentration $r$ of a chemical via
membrane receptors binding the chemical's molecules
\cite{Adler69,Lacki81}. It is crucial to note that when estimating
gradients of chemicals, the cells actually measure changes in the
receptors' occupancy and not in the concentration itself. When put in
continuous equations \cite{Murray89,GKCB98}, this indirect
measurement translates to measuring the gradient
\begin{equation}
\frac \partial {\partial x}\frac r{\left( K+r\right) }=\frac
K{(K+r)^2}\frac{\partial r}{\partial x}.
\label{eq:bioBG:receptorLow}
\end{equation}
where $K$ is a constant whose value depends on the receptors'
affinity, the speed in which the bacterium processes the signal from
the receptor, etc. This means that the chemical gradient times a
factor $K/(K+r)^2$ is measured, and it is known as the ``receptor
law'' \cite{Murray89}.

When modeling chemotaxis performed by walkers, it is possible to
modulate the periods between tumbling (without changing the speed) in
the same way the bacteria do. It can be shown that step length
modulation has the same mean effect as keeping the step length
constant and biasing the direction of the steps (higher probability
to move in the preferred direction). As this later approach is
numerically simpler, this is the one implemented in the Communicating
Walkers model.

In a continuous model, we incorporate the effect of chemotaxis by
introducing a chemotactic flux $\vec{J}_{chem}$:
\begin{equation}
\vec{J}_{chem}\equiv \zeta (\sigma )\chi (r)\nabla r
\label{j_chem}
\end{equation}
$\chi (r)\nabla r$ is the gradient sensed by the cell (with $\chi
(r)$ having the units of 1 over chemical's concentration). $\chi (r)$
is usually taken to be either constant or the ``receptor law''.
$\zeta (\sigma )$ is the bacterial response to the sensed gradient
(having the same units as a diffusion coefficient). In the Non-Linear
Diffusion model the bacterial diffusion is $D_b=D_0b^k$, and the
bacterial response to chemotaxis is $\zeta (b)=\zeta _0b\left(
D_0b^k\right) =\zeta _0D_0b^{k+1}$. $\zeta _0$ is a constant,
positive for attractive chemotaxis and negative for repulsive
chemotaxis.

Ben-Jacob \etal argued \cite{BCC96,CCB96,BC97,BenJacob97} that for
the colonial adaptive self-organization the bacteria employ three
kinds of chemotactic responses, each dominant in different regime of
the morphology diagram. One response is the food chemotaxis mentioned
above. It is expected to be dominant for only a range of nutrient
levels (see the ``receptor law'' below). The two other kinds of
chemotactic responses are signaling chemotaxis. One is long-range
repulsive chemotaxis. The repelling chemical is secreted by starved
bacteria at the inner parts of the colony. The second signal is a
short-range attractive chemotaxis.
The length scale of each signal is determined by the diffusion
constant of the chemical agent and the rate of its spontaneous
decomposition.

{\em Amplification of diffusive Instability Due to Nutrients
Chemotaxis:}
In non-living systems, more ramified patterns (lower fractal
dimension) are observed for lower growth velocity. Based on growth
velocity as function of nutrient level and based on growth dynamics,
Ben-Jacob \etal \cite {BSTCCV94a} concluded that in the case of
bacterial colonies there is a need for mechanism that can both
increase the growth velocity and maintain, or even decrease, the
fractal dimension. They suggested food chemotaxis to be the required
mechanism. It provides an outward drift to the cellular movements;
thus, it should increase the rate of envelope propagation. At the
same time, being a response to an external field it should also
amplify the basic diffusion instability of the nutrient field. Hence,
it can support faster growth velocity together with a ramified
pattern of low fractal dimension.

The above hypothesis was tested in the Communicating Walkers model
and in the Non-Linear Diffusion model. In figures 9 and 10 it is
shown that as expected, the inclusion of food chemotaxis in both
models led to a considerable increase of the growth velocity without
significant change in the fractal dimension of the pattern.

{\em Repulsive chemotactic signaling: }
We focus now on the formation of the fine radial branching patterns
at low nutrient levels. From the study of non-living systems, it is
known that in the same manner that an external diffusion field leads
to the diffusion instability, an internal diffusion field will
stabilize the growth. It is natural to assume that some sort of
chemotactic agent produces such a field. To regulate the organization
of the branches, it must be a long-range signal. To result in radial
branches it must be a repulsive chemical produced by cells at the
inner parts of the colony. The most probable candidates are the
bacteria entering a pre-spore stage.
\label{sec:repulsive}

If nutrient is deficient for a long enough time, bacterial cells may
enter a special stationary state -- a state of a spore -- which
enables them to survive much longer without food.
While the spores themselves do not emit any chemicals (as they have
no metabolism), the pre-spores (sporulating cells) do not move and
emit a very wide range of waste materials, some of which unique to
the sporulating cell. These emitted chemicals might be used by other
cells as a signal carrying information about the conditions at the
location of the pre-spores. Ben-Jacob \etal \cite
{BSTCCV94a,BSTCCV94b,CCB96} suggested that such materials are
repelling the bacteria ('repulsive chemotactic signaling') as if they
escape a dangerous location.

The equation describing the dynamics of the chemorepellent contains
terms for diffusion, production by pre-spores, decomposition by
active bacteria and spontaneous decomposition:
\begin{equation}
\frac{\partial r}{\partial t}=D_r{\nabla ^2 r+\Gamma }_r s-\Omega _r
b
r- \lambda_r r
\label{r-eqn}
\end{equation}
where $D_r$ is the diffusion coefficient of the chemorepellent,
$\Gamma _r$ is the emission rate of repellent by pre-spores, $\Omega
_r$ is the decomposition rate of the repellent by active bacteria,
and $\lambda _r$ is the rate of self decomposition of the repellent.
In the Communicating Walkers model $b$ and $s$ are replaced by active
and inactive walkers, respectively.

In figures 9 and 10 the effect of repulsive chemotactic signaling is
shown. In the presence of repulsive chemotaxis the patterns in both
models become much denser with a smooth circular envelope, while the
branches are thinner and radially oriented.

\section{Conclusions} 

We show here a pattern forming system, bacterial colony, whose 
discrete elements, the bacteria, are big enough to raise the question 
of modeling discrete systems. We study two types of models. The 
Communicating Walkers model has explicit discrete units to represent 
the bacteria. The ratio between the walkers' size and the pattern's 
size is even bigger than the ratio in the bacterial colony. The 
second type of models is continuous reaction-diffusion equations. 
Non-linear diffusion causes a sharp boundary to appear in these 
models. Following Kessler and Levine \cite{KL98}, we account for the 
discreteness of the bacteria by including a cutoff in the bacterial 
growth term. The cutoff does not improve the models' descriptive 
power. The main effect of such cutoff is to decrease the width of the 
colony's front, making the growth pattern more sensitive to effects 
such as implicit anisotropy. We conclude that the presence of a 
boundary cancels the need for explicit treatment of discreteness.

In order to assess the similarity between the discrete Communicating 
Walkers model and the continuous Non-Linear Diffusion model, we 
incorporate food chemotaxis and repulsive chemotactic signaling into 
the models (both are expected to exist in the bacterial colonies). 
Both models respond to such changes in the same way, exhibiting 
altered patterns and altered dynamics, similar to those observed in 
the bacterial colonies. From this similarity we conclude that to some 
extent inferences from one model can be applied to the other. 
Specifically we focus on insensitivity of the Communicating Walkers 
model to implicit anisotropy and on the sensitivity a cutoff imposes 
on the continuous models. From the two facts combined we conclude 
that the magnified discreteness in the Communicating Walkers model is 
still small enough to be neglected. 

\medskip \noindent {\it Acknowledgements}. 
We have benefited from many discussions on the presented studies with 
H. Levine. IG wishes to thank R. Segev for fruitful discussions. 
Identifications of the \Tname* and genetic studies are carried in 
collaboration with the group of D. Gutnick. Presented studies are 
supported in part by a grant from the Israeli Academy of Sciences 
grant no. 593/95 and by the Israeli-US Binational Science Foundation 
BSF grant no. 00410-95.

\section*{Figure Captions}

\begin{figure}[hbp] 
% \centerline{ 
% \psfig{figure=figure1.ps,width=4.7in} } 
\makecaption {Fig.~1}
{ 
Observed branching patterns of colonies of \Tname* grown on 2\% agar  
concentration. The nutrient level is, from left to right, 0.25 gram  
peptone per liter, 0.5$g/l$ and 5$g/l$. The colony on the right has  
wide branches, much wider than the gaps between them, can be seen.  
The pattern in the middle is less ordered, fractal-like pattern,  
similar to patterns seen in electro-chemical deposition and DLA  
simulations \protect\cite{BG90,BenJacob93}. As the nutrient level is further  
decreased the pattern become denser again, with pronounced circular  
envelope (on the left). 
\label{fig:T:colonies} }
\end{figure} 

\begin{figure}[hbp] 
% \centerline{\psfig{figure=figure2a.ps,height=4cm} ~~~  
% \psfig{figure=figure2b.ps,height=4cm} } 
\makecaption{Fig.~2}{ 
Closer look on a branch of a bacterial colony. The left figure shows  
the lubrication fluid in which the bacteria are immersed. On the  
right, the individual bacteria can be seen. Each dot in the branch is  
a $1\times2 \mu m$ bacterium. The dots outside the branch are not  
bacteria but deformations of the agar. 
\label{fig:T:branch}} 
\end{figure} 

\begin{figure}[hbp] 
% \centerline{\psfig{figure=figure3.ps,width=4.7in} }  
\makecaption{Fig.~3}{ 
Colonial pattern of the Communicating Walkers model. 
Here $N_c=20$ and $n_0$ is 6, 8, 10 and 30 from left to right  
respectively.  
\label{fig:CW:colonies}} 
\end{figure} 

\begin{figure}[hbp] 
% \centerline{ 
% \psfig{figure=figure4a.ps,height=4cm} ~~~  
% \psfig{figure=figure4b.ps,height=4cm} } 
\makecaption{Fig.~4}{ 
Closer look on simulated colonies. On the right: a tip of a branch in  
the Communicating Walkers model. The boundary of the branch and 
walkers can be seen. On the left: lubrication at a tip of a branch in 
the Lubricating Bacteria model. 
\label{fig:models:branches}} 
\end{figure} 

\begin{figure}[hbp] 
% \centerline{ 
% \psfig{figure=figure5.ps,width=4.7in} }  
\makecaption{Fig.~5}{ 
Growth patterns of the Lubricating Bacteria model, for different 
values of initial nutrient level $n_0$. The apparent (though weak) 6-
fold anisotropy is due to the underlying tridiagonal lattice. 
\label{fig:lubrication:pepton}} 
\end{figure} 

\begin{figure}[hbp] 
% \centerline{ 
% \psfig{figure=figure6.ps,width=4.7in} }  
\makecaption{Fig.~6}{ 
The effect of a cutoff on the growth patterns in the
Lubricating Bacteria model. Aside from the cutoff, the conditions are
the same as 
in the middle pattern of figure 5, where the maximal value of $b$ was 
about 0.025. The values of the cutoff $\beta $ are, from left to 
right, $ 10^{-6}$, $10^{-5}$ and $3\cdot 10^{-5}$. The 6-fold 
symmetry is due to anisotropy of the underlying lattice which is 
enhanced by the cutoff. 
\label{fig:lubrication:cutoff}} 
\end{figure} 

\begin{figure}[hbp]
% \centerline{
% \psfig{figure=figure7.ps,width=4.7in} }
\makecaption{Fig.~7}{
Growth patterns of the Kitsunezaki model, for different
values of initial nutrient level $n_0$. Parameters are:  $D_0=0.1, k=1, \mu=0.15$.
The apparent 6-fold
symmetry is due to the underlying tridiagonal lattice.
\label{fig:k1}}
\end{figure}

\begin{figure}[hbp]
% \centerline{ \psfig{figure=figure8.ps,height=4cm} }
\makecaption{Fig.~8}{
Growth patterns of the Kitsunezaki model, with a cutoff correction. 
Cutoff value $\beta=0.1$, all other parameters as in Fig. 7, right
pattern.
The apparent 6-fold symmetry is due to the underlying tridiagonal
lattice.
\label{fig:k-cut}}
\end{figure}

\begin{figure}[hbp]
% \centerline{
% \psfig{figure=figure9a.ps,height=4cm} ~~~
% \psfig{figure=figure9b.ps,height=4cm} }
\makecaption{Fig.~9}{
The effect of chemotaxis on growth in the Communicating Walkers
model. On the left: chemotaxis towards food is added to the model.
The conditions are the same as in figure 3, second from right
pattern. The pattern is essentially unchanged by food chemotaxis, but
the growth velocity is almost doubled. On the right: repulsive
chemotactic signaling is added to the model. The conditions are the
same as in figure 3, left pattern. The pattern is of fine radial
branches with circular envelope, like in figure 1, left pattern.
\label{fig:CW:chemotaxis}}
\end{figure}

\begin{figure}[hbp]
% \centerline{ \psfig{figure=figure10.ps,height=4cm} }
\makecaption{Fig.~10}{
Growth patterns of the Non-Linear Diffusion model with food
chemotaxis (left) and repulsive chemotactic signaling (right)
included. $\chi _{0f}=3,\chi _{0r}=1,D_r=1,\Gamma _r=0.25,\Omega
_r=0,\Lambda _r=0.001$. Other parameters are the same as in figure 7.
The apparent 6-fold symmetry is due to the underlying tridiagonal
lattice.
\label{fig:k-food-rep}}
\end{figure}


\section*{References}


\begin{thebibliography}{10}

\bibitem{Kepler}
J.~Kepler.
\newblock {\em De Nive Sexangula Godfrey Tampach}.
\newblock Frankfurt am Main, 1611.

\bibitem{Mandelbrot77b}
B.B. Mandelbrot.
\newblock {\em The Fractal Geometry of Nature}.
\newblock Freeman, San Francisco, 1977.

\bibitem{Vicsek89}
T.~Vicsek.
\newblock {\em Fractal Growth Phenomena}.
\newblock World Scientific, New York, 1989.

\bibitem{BenJacob93}
E.~{Ben-Jacob}.
\newblock {\em Contemp. Phys.} {\bf 34} (1993) 247.

\bibitem{BenJacob97}
E.~{Ben-Jacob}.
\newblock {\em Contemp. Phys.} {\bf 38} (1997) 205.

\bibitem{KKL88}
D.~A. Kessler, J.~Koplik, and H.~Levine.
\newblock {\em Adv. Phys.} {\bf 37} (1988) 255.

\bibitem{Langer89}
J.S. Langer.
\newblock {\em Science} {\bf 243} (1989) 1150.

\bibitem{BG90}
E.~{Ben-Jacob} and P.~Garik.
\newblock {\em Nature} {\bf 343} (1990) 523.

\bibitem{BCL98}
E.~{Ben-Jacob}, I.~Cohen, and H.~Levine.
\newblock Cooperative self-organization of microorganisms.
\newblock {\em Adv. Phys.},  in press.
\newblock (1998).

\bibitem{BSST92}
E.~{Ben-Jacob}, H.~Shmueli, O.~Shochet, and A.~Tenenbaum.
\newblock {\em Physica A} {\bf 187} (1992) 378.

\bibitem{BTSA94}
E.~{Ben-Jacob}, A.~Tenenbaum, O.~Shochet, and O.~Avidan.
\newblock {\em Physica A} {\bf 202} (1994) 1.

\bibitem{TBG98}
M.~Tcherpikov, E.~{Ben-Jacob}, and D.~Gutnick.
\newblock Identification of two pattern-forming strains and their localization
  in a phylogenetic cluster.
\newblock {\em Int. J. Syst. Bacteriol.},  in press.
\newblock (1998).

\bibitem{FM89}
H.~Fujikawa and M.~Matsushita.
\newblock {\em J. Phys. Soc. Jap.} {\bf 58} (1989) 3875.

\bibitem{MF90}
M.~Matsushita and H.~Fujikawa.
\newblock {\em Physica A} {\bf 168} (1990) 498.

\bibitem{MKNIHY92}
Matsuyama T; Kaneda K; Nakagawa Y; Isa K; H. Hara-Hotta;~Yano I.
\newblock {\em J. Bacteriol.} {\bf 174} (1992) 1769.

\bibitem{MM95}
T.~Matsuyama and M.~Matsushita.
\newblock In P.~M. Iannaccone and M.~K. Khokha, editors, {\em Farctal Geometry
  in Biological Systems, an Analytical Approach}, pages 127-171. CRC Press,
  New-York, 1995.

\bibitem{MS96}
N.~H. Mendelson and B.~Salhi.
\newblock {\em J. Bacteriol.} {\bf 178} (1996) 1980.

\bibitem{KL98}
D. A. Kessler and H. Levine,''Fluctuation-induced diffusive instabilities'',
  Nature, in press.

\bibitem{BSTCCV94a}
E.~{Ben-Jacob}, O.~Shochet, A.~Tenenbaum, I.~Cohen, A.~Czir\'ok, and T.~Vicsek.
\newblock {\em Nature} {\bf 368} (1994) 46.

\bibitem{SKBLM92a}
O.~Shochet, K.~Kassner, E.~{Ben-Jacob}, S.G. Lipson, and
  H.~{M\"{u}ller-Krumbhaar}.
\newblock {\em Physica A} {\bf 181} (1992) 136.

\bibitem{SKBLM92b}
O.~Shochet, K.~Kassner, E.~{Ben-Jacob}, S.G. Lipson, and
  H.~{M\"{u}ller-Krumbhaar}.
\newblock {\em Physica A} {\bf 187} (1992) 87.

\bibitem{Shochet95}
O.~Shochet.
\newblock {\em Study of late-stage growth and morphology selection during
  diffusive patterning}.
\newblock PhD thesis, Tel-Aviv University, 1995.

\bibitem{BCC95b}
E.~{Ben-Jacob}, I.~Cohen, and A.~Czir\'ok.
\newblock {\em Fractals} {\bf } (1997) .
\newblock (in press).

\bibitem{BCC96}
E.~{Ben-Jacob}, I.~Cohen, and A.~Czir\'ok.
\newblock Smart bacterial colonies.
\newblock In {\em Physics of Biological Systems: From Molecules to Species},
  Lecture Notes in Physics, pages 307-324. Springer-Verlag, Berlin, 1997.

\bibitem{CCB96}
I.~Cohen, A.~Czir\'ok, and E.~{Ben-Jacob}.
\newblock {\em Physica A} {\bf 233} (1996) 678.

\bibitem{BC97}
E.~{Ben-Jacob} and I.~Cohen.
\newblock Cooperative formation of bacterial patterns.
\newblock In J.~A. Shapiro and M.~Dworkin, editors, {\em Bacteria as
  Multicellular Organisms}. Oxford University Press, New-York, 1997.

\bibitem{GKCB98}
I.~Golding, Y.~Kozlovsky, I.~Cohen, and E.~{Ben-Jacob}.
\newblock Studies of bacterial branching growth using reaction-diffusion models
  of colonial development.
\newblock {\em Physica A},  in press, cond-mat/9807088.
\newblock (1998).

\bibitem{Cohen97}
I. Cohen, {\em Mathematical Modeling and Analysis of Pattern Formation and
  Colonial Organization in Bacterial Colonies}, MSc thesis, Tel-Aviv
  University, ISRAEL, 1997.

\bibitem{Kitsunezaki97}
S.~Kitsunezaki.
\newblock {\em J. Phys. Soc. Jpn} {\bf 66}(5) (1997) 1544.

\bibitem{Adler69}
J.~Adler.
\newblock {\em Science} {\bf 166} (1969) 1588.

\bibitem{BP77}
H.~C. Berg and E.~M. Purcell.
\newblock {\em Biophysical Journal} {\bf 20} (1977) 193.

\bibitem{Lacki81}
J.~M. Lackiie, editor.
\newblock {\em Biology of the chemotactic response}.
\newblock Cambridge Univ. Press, 1986.

\bibitem{Berg93}
H.~C. Berg.
\newblock {\em Random Walks in Biology}.
\newblock Princeton University Press, Princeton, N.J., 1993.

\bibitem{Murray89}
J.~D. Murray.
\newblock {\em Mathematical Biology}.
\newblock Springer-Verlag, Berlin, 1989.

\bibitem{BSTCCV94b}
E.~{Ben-Jacob}, O.~Shochet, A.~Tenenbaum, I.~Cohen, A.~Czir\'ok, and T.~Vicsek.
\newblock {\em Fractals} {\bf 2}(1) (1994) 15.

\end{thebibliography}
\end{document}